\documentclass[10pt,letterpaper]{article}
\usepackage[left=0.75in, right=0.75in, top=0.75in, bottom=1.00in]{geometry}
\usepackage{mathptmx}				
\usepackage[hidelinks]{hyperref}	
\usepackage[english]{babel}			
\usepackage{blindtext}				
\usepackage{graphicx}				
\usepackage{booktabs}				
\usepackage{caption}				
\usepackage{float}					
\usepackage{amsmath}				
\usepackage{amsfonts}				
\usepackage{cite}					
\usepackage{nopageno} 				
\usepackage{gensymb}

\makeatletter
\newcommand{\removeperiod}{\@ifnextchar.{\@gobble}\relax}		
\makeatother

\usepackage{etoolbox}
\makeatletter
\patchcmd{\@maketitle}{\Large}{\Large}{\typeout{OK 1}}{\typeout{Failed 1}}
\patchcmd{\@maketitle}{\large \lineskip}{\large \lineskip}{\typeout{OK 2}}{\typeout{Failed 2}}
\makeatother

\usepackage[explicit]{titlesec}
\renewcommand{\thesection}{}
\titleformat{\section}
  {\normalfont}{\thesection}{0pt}{\MakeUppercase{\textbf{#1}}}
\titlespacing*{\section}
  {0pt}{5ex}{-1em}

\newcommand{\keywords}[1]{\vspace{2ex} \noindent \textbf{Keywords:} #1}

\begin{document}
	\date{} 
	
	\title{Seal Whisker-Inspired Sensor for Amplifying Wake-Induced Vibrations in Underwater Marine Animal Monitoring}
	
	\author{\vspace{.25in}\\
	Yuyan Wu$^1$, Sanjay Giridharan$^2$, Leixin Ma$^2$ and Hae Young Noh$^1$\\ \\ 
	 $^1$Department of Civil Engineering \\
		Stanford University, California, CA 94305\\ \\
$^2$School for Engineering of Matter, Transport and Energy \\
		Arizona State University, Tempe, AZ 85287
	}
	\maketitle
	\section{abstract}
    Underwater marine animal monitoring is essential for assessing biodiversity, evaluating ecosystem health, and understanding the effects of offshore structures. Traditional approaches such as tagging, sonar, and camera systems are often invasive, energy-intensive, or limited by poor visibility and water turbidity. Inspired by the hydrodynamic sensing of seal whiskers, wavy whisker vibration sensors have been developed for flow velocity and angle-of-attack detection. However, most prior work has focused on sensor characterization and only forward modeling, with limited exploration of the inverse problem of inferring animal movement. Moreover, current sensor sensitivity to vortex street wakes generated by swimming animals remains insufficient for practical monitoring. To address this gap, we develop a whisker-inspired sensor with a spiral-perforated base that amplifies vibrations within frequency ranges relevant to animal-induced wakes. We further characterize the influence of spiral parameters on the sensitive frequency band, enabling adaptation of the design to specific species. We evaluated the amplification effect of the spiral-perforated design using frequency response simulations of the whisker-base structure under harmonic water pressure. Results show up to 51× enhancement in root mean squared displacement at the target sensor location within frequency bands associated with animal-induced wakes compared to the baseline design, confirming the effectiveness of the amplification.
		
	\keywords{Underwater Marine Animal Monitoring; Bio-inspired Sensing; Seal Whisker; Vibration Amplification Structure; Wake-Induced Vibrations}

\section{Introduction}
Underwater marine animal monitoring, including tracking their populations and behaviors, is essential for understanding biodiversity, assessing ecosystem health, and evaluating the ecological impacts of offshore infrastructure. Keystone species~\cite{paine1969note} such as whales and sharks play critical roles in maintaining the underwater food web balance~\cite{roman2014whales, heithaus2008predicting, dedman2024ecological}. Monitoring their populations is important for detecting early signs of ecological imbalance. In addition, many marine animals, such as corals, fish, and mammals, are sensitive to temperature shifts, ocean acidification, and oxygen depletion~\cite{doney2020impacts, kroeker2013impacts, alter2024hidden}. Monitoring changes in their distribution, migration, and behavior provides critical insight into the effects of climate change on marine ecosystems. Furthermore, marine animal monitoring is crucial for understanding how offshore infrastructure, such as wind farms and oil rigs, affects marine fauna. Construction and operational noise can disrupt communication, navigation, and foraging~\cite{brandt2009harbour,tougaard2009pile}. Infrastructure may also create artificial reefs that attract certain species or disrupt migration routes of sea turtles, fish, and marine mammals~\cite{langhamer2012artificial,barnette2017potential}. Animal monitoring identifies ecological changes and reveals how animals adapt to offshore infrastructures, enabling governments to implement adaptive management strategies that minimize harm.


Existing methods for monitoring marine animals include tagging, camera-based systems, acoustic sensing, and artificial lateral lines (ALL). Tagging involves attaching archival biologgers or transmitting biotelemetry to record individual movement and behavior~\cite{jepsen2015use, pine2012design}. This method provides detailed, long-term data but is invasive and impractical for large populations. Cameras offer a non-invasive alternative, capturing images to monitor species population and behavior~\cite{ruff1995fish,boom2012long,chuang2016underwater}. However, its effectiveness is constrained by underwater lighting conditions, making it unsuitable in dark or turbid environments. Acoustic sensing, mainly including sonar, transmits sound pulses and analyzes returning echoes to track the population and movement of marine animals and identify species~\cite{hozyn2021review,hodges2011underwater,marage2013sonar}. It is non-invasive, effective in turbid or dark waters, and covers large ranges. However, its active sensing mechanism is energy-intensive. ALL systems, inspired by the sensory organs of fish and amphibians, detect flow velocity and pressure gradients through arrays of pressure or flow sensors (e.g., MEMS-based or piezoelectric) mounted on underwater robots~\cite{fan2002design, kottapalli2014touch, venturelli2012hydrodynamic}. They are non-invasive, energy-efficient, and robust in dark or turbid waters. However, ALL sensors exhibit limited sensitivity to wakes generated by marine animals and are highly susceptible to turbulence and background currents in complex underwater environments, which reduces their effectiveness for long-distance detection~\cite{chambers2014fish, liu2016review}. 

To overcome these limitations, we introduce seal whisker-inspired sensors for underwater marine animal monitoring. These sensors emulate the mechanism by which seals detect prey using their whiskers. Swimming marine animals generate vortex street wakes through body movement, and these wakes induce vibrations in seal whiskers at their characteristic natural frequency. Seals then detect these vibrations to localize and track prey~\cite{zheng2021creating}. Seal whiskers have a wavy morphology to suppress self-induced vortices and thereby enhance the signal-to-noise ratio for long-distance wake detection~\cite{dehnhardt2001hydrodynamic, hanke2010harbor}. Compared to other methods, whisker-inspired sensing is non-invasive, passive, energy-efficient, and independent of lighting conditions. By reducing self-motion noise, it enables more reliable detection of animal-induced vortex street wakes over long distances. Seal whisker-inspired sensors have recently been used to measure hydrodynamic parameters, such as flow velocity, angle of attack, and pressure~\cite{zhang2021harbor,dai2024biomimetic, zheng20223d, beem2012calibration}. 
However, existing whisker-inspired sensor works primarily focus on characterization and only forward modeling, using wavy morphologies to suppress self-induced vortices~\cite{kottapalli2015harbor, dai2024biomimetic, zheng20223d, beem2012calibration}. The inverse problem of inferring marine animal movement remains challenging, as animal-induced wakes are small in amplitude and difficult to detect. Swimming marine animals typically induce velocity fluctuations on the order of several centimeters per second~\cite{nauen2002quantification, drucker2001wake}, corresponding to dynamic pressure variations of roughly 1 Pa (according to the Bernoulli relation). This magnitude falls below the sensitivity threshold of current underwater pressure-sensing arrays~\cite{yaul2012flexible}. Furthermore, these vortices dissipate rapidly~\cite{hanke2004hydrodynamic}, making detection even more challenging. 


We address this challenge through a novel sensor hardware design of a spiral perforated base that amplifies whisker vibrations within targeted frequency ranges. By adjusting the base geometry, the design can be tuned to amplify whiskers with different natural frequencies, enabling customization for specific target frequency bands. This amplification enhances the signal-to-noise ratio for underwater marine animal monitoring by amplifying the desired vibration response while suppressing noise from self-motion and background underwater disturbances with a wavy shape whisker geometry. Furthermore, we characterize how geometric parameters of the whisker base influence amplification rates and frequency-band shifts. This characterization provides guidelines for tuning geometry parameters to optimize amplification for different target frequency bands.

We evaluated the effectiveness of the whisker base design by conducting frequency-response simulations and comparing the results with a conventional whisker sensor design. The simulations demonstrate that our whisker base design amplifies whisker vibration amplitude by up to 51 times.

The main contributions of this paper are as follows:
\begin{itemize}

\item We develop a seal-whisker-inspired sensor for underwater marine animal monitoring through whisker vibrations induced by animal-generated wakes.

\item To overcome the challenge of inherently small amplitudes of animal-induced wakes, we design a novel sensor hardware with a spiral-perforated whisker base that amplifies wake-induced vibrations within the target frequency range, thereby enhancing the sensitivity of the whisker-inspired sensor. We further characterize how whisker geometry parameters influence amplification rates and frequency bands.

\item We evaluated the proposed design through frequency-response simulations of the seal whisker–base system, demonstrating the effectiveness of amplification.
\end{itemize}

In the remainder of this paper, we first introduce the background on the mechanisms by which seals detect prey through their whiskers and the geometric characteristics of seal whiskers. We then present the design of the whisker base and characterize how the geometric parameters of the whisker base affect amplification and the associated frequency ranges. Finally, we evaluate the effectiveness of amplification through frequency-response simulations. 

\section{Background: Mechanisms of Animal Detection by Seal Whiskers}


\begin{figure}[t]
    \centering
    \includegraphics[width=\linewidth]{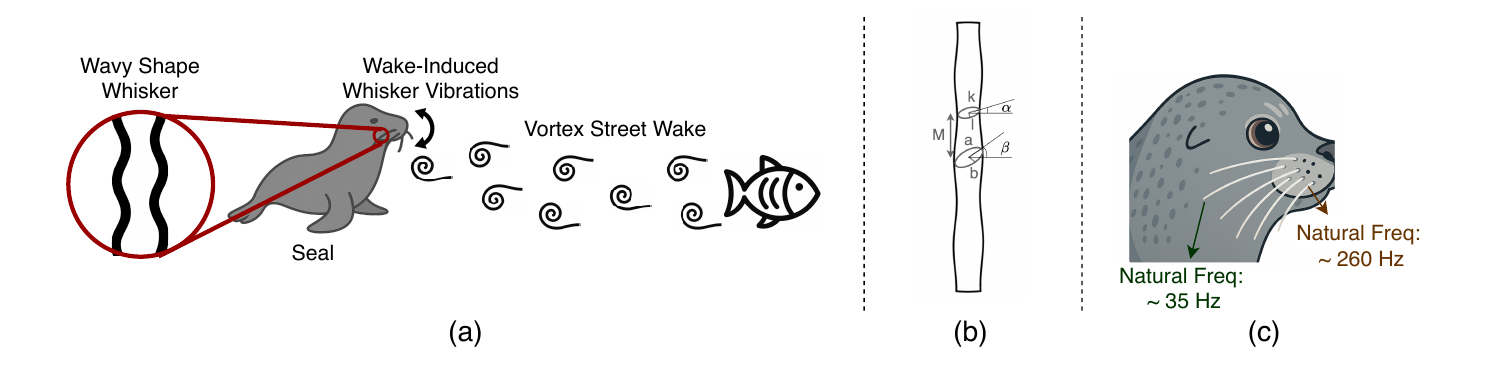}
    \caption{(a) Seals track prey using wavy shape whiskers that detect vibrations induced by vortex street wakes generated by swimming animals. (b) Enlarged view of the wavy whisker cross-section, illustrating elliptical geometry with major and minor axes $(a, b, k, l)$ and orientation angles $(\alpha, \beta)$. (c) Seals have an array of whiskers with various natural frequencies, facilitating broadband hydrodynamic sensing across multiple frequency ranges.}
    \label{fig:intuition}
\end{figure}


Seals can detect prey over 180 meters away using their whiskers~\cite{kamat2024undulating}. As marine animals move underwater, they generate trailing vortices known as reverse von Kármán vortex streets~\cite{triantafyllou2000hydrodynamics}. When these vortex street wakes pass over seal whiskers, the whiskers vibrate in synchrony with the wakes (see Figure~\ref{fig:intuition}a). Mechanoreceptors at the whisker base transduce these vibrations into neural signals that provide information about prey location and movement. Moreover, because different marine animals produce distinct vortex patterns~\cite{tytell2004hydrodynamics, sfakiotakis2002review}, seals can also identify prey species from the wake-induced whisker vibrations~\cite{wieskotten2011hydrodynamic}. Seal whiskers exhibit a characteristic wavy geometry with elliptical cross-sections (see Figure~\ref{fig:intuition}b). This geometry can be parameterized by $(a,b,k,l,M,\alpha,\beta)$ as defined in Figure~\ref{fig:intuition}b, where $(a,b)$ and $(k,l)$ the lengths of the semi-major and semi-minor axes of the maximum and minimum elliptical cross-sections, respectively; $M$ denotes the distance between the adjacent cross-sections; and $(\alpha, \beta)$ specify the orientation angles of these cross-sections. The whisker tapers from a thick base to a gradually thinner tip. Furthermore, seals possess whiskers with diverse natural frequencies, enabling detection across a broad frequency band (Figure~\ref{fig:intuition}c)~\cite{zheng2025wonders}.


Seal whisker-inspired sensors~\cite{dai2024biomimetic, zheng20223d, beem2012calibration,zhang2021harbor} replicate the wavy geometry of seal whiskers, which suppress self-induced vibrations during swimming and filter out background noise~\cite{dehnhardt2001hydrodynamic, hanke2010harbor}. When a straight, cylindrical whisker moves through water, it creates strong von Kármán vortex streets, which are the alternating vortices that are shed off the whisker in a regular pattern~\cite{morkovin1964flow,beaudan1995numerical}. This vortex shedding produces vibrations (self-noise) that mask the tiny water movements caused by prey. The wavy geometry disrupts these regular vortex streets, breaking them into smaller, less coherent vortices~\cite{dehnhardt2001hydrodynamic, hanke2010harbor}. This reduces the strength of vibrations transmitted to the whisker base, making them less sensitive to noise induced by self-motion. These wavy shape whiskers enable seals and whisker-inspired sensors to detect and track hydrodynamic wakes left by animals over a long distance. However, current studies on whisker-inspired sensors have primarily focused on mitigating self-induced vortices~\cite{kottapalli2015harbor,dai2024biomimetic, zheng20223d, beem2012calibration}, with limited exploration of mechanisms to enhance sensitivity to animal-induced vortices.

\section{Whisker-Inspired Sensor Design for Wake-Induced Vibration Amplification}

\begin{figure}[htbp]
    \centering
    \includegraphics[width=0.6\linewidth]{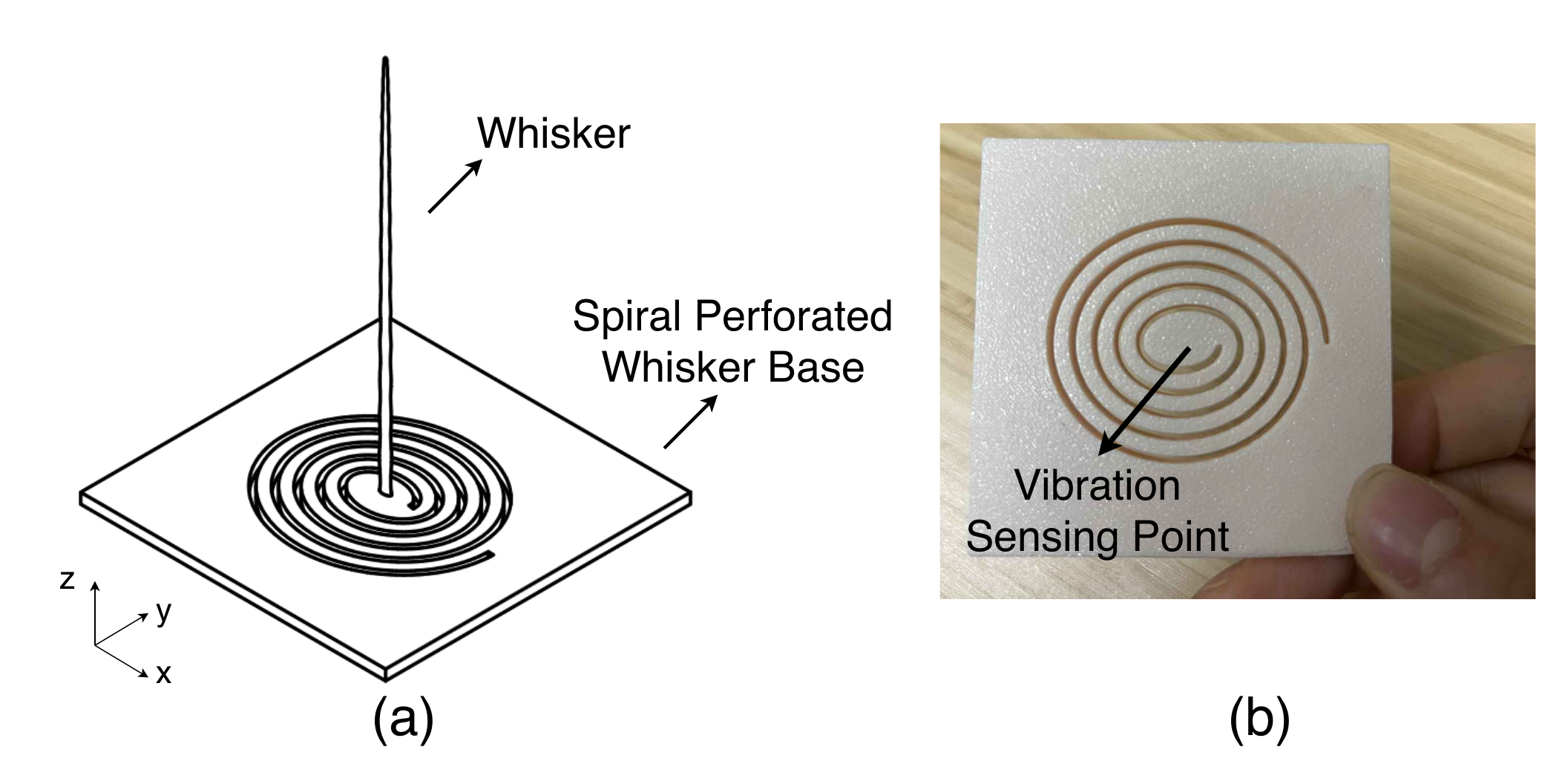}
    \caption{(a) Design of the whisker-inspired sensing unit and spiral perforated base. The whisker is mounted on a spiral perforated base, which amplifies vortex-induced vibrations. (b) Photo of the spiral perforated base. The vibration sensing point is at the middle of the base, where a vibration sensor will be mounted to capture vortex-induced whisker vibrations.}
    \label{fig:base_design}
\end{figure}

To enhance sensitivity to wake-induced vibrations within the frequency range of the animal-induced wakes, we designed a spiral-perforated whisker base that increases structural flexibility and amplifies vibration amplitude in the target band. The whisker is mounted at the base center (Figure~\ref{fig:base_design}a), with the bottom view shown in Figure~\ref{fig:base_design}b. The sensing point, located on the opposite surface directly beneath the whisker attachment, is intended for embedding a vibration sensor. This spiral-perforated design amplifies displacement at the sensing point under harmonic pressures generated by animal-induced wakes. 

The spiral perforations reduce the effective stiffness of the whisker base by extended deformation paths, thereby amplifying the vibrations of the whiskers. Modeling the whisker base system as a single degree-of-freedom (SDOF) translational oscillator, its displacement $x(t)$ under vortex forcing $F_0e^{j\omega t}$ is governed by
\begin{equation}
    m\ddot{x}(t)+c\dot{x}(t)+kx(t) = F_0e^{j\omega t},
\end{equation}
where $m$ is the combined effective mass of the whisker and base, $c$ is the viscous damping coefficient, and $k$ is the effective stiffness determined by the spiral-perforated base and whisker structure. The steady-state harmonic response is
\begin{equation}
    X(\omega) = \frac{F_0}{k-m\omega^2+jc\omega},\ \  |X(\omega)| = \frac{F_0}{\sqrt{(k-m\omega^2)^2+(c\omega)^2}}
    \label{eq:x}
\end{equation}
with $X(\omega)$ denoting the Fourier-domain displacement amplitude. Modeling the solid segments within the spiral region of the base as beam elements, the effective stiffness $k\propto \frac{EI}{L^3}$, where $E$ is the Young’s modulus, $I=\frac{bh^3}{12}$ is the second moment of area with $b$ representing the width of the spiral region and $h$ the thickness of the whisker base, and $L$ is the effective load-path length, positively correlated to the total spiral beam length. The spiral geometry increases $L$ and reduces $b$, thereby decreasing $k$ and enhancing the vibration amplitude $|X(\omega)|$.

Vibration amplification occurs through resonance coupling when the whisker and base exhibit closely matched natural frequencies in a coupled vibration mode. In the case where the exciter frequency lies below the natural frequency of the coupled system, the whisker and base primarily respond in phase. Under this same-phase condition, resonance coupling reinforces their motion and amplifies vibration, enabling efficient energy transfer~\cite{den1985mechanical,rana1998parametric}. At resonance, the whisker and base interact strongly, oscillate coherently, and enhance the efficiency of energy transmission.

\begin{figure}[htbp]
    \centering
    \includegraphics[width=0.6\linewidth]{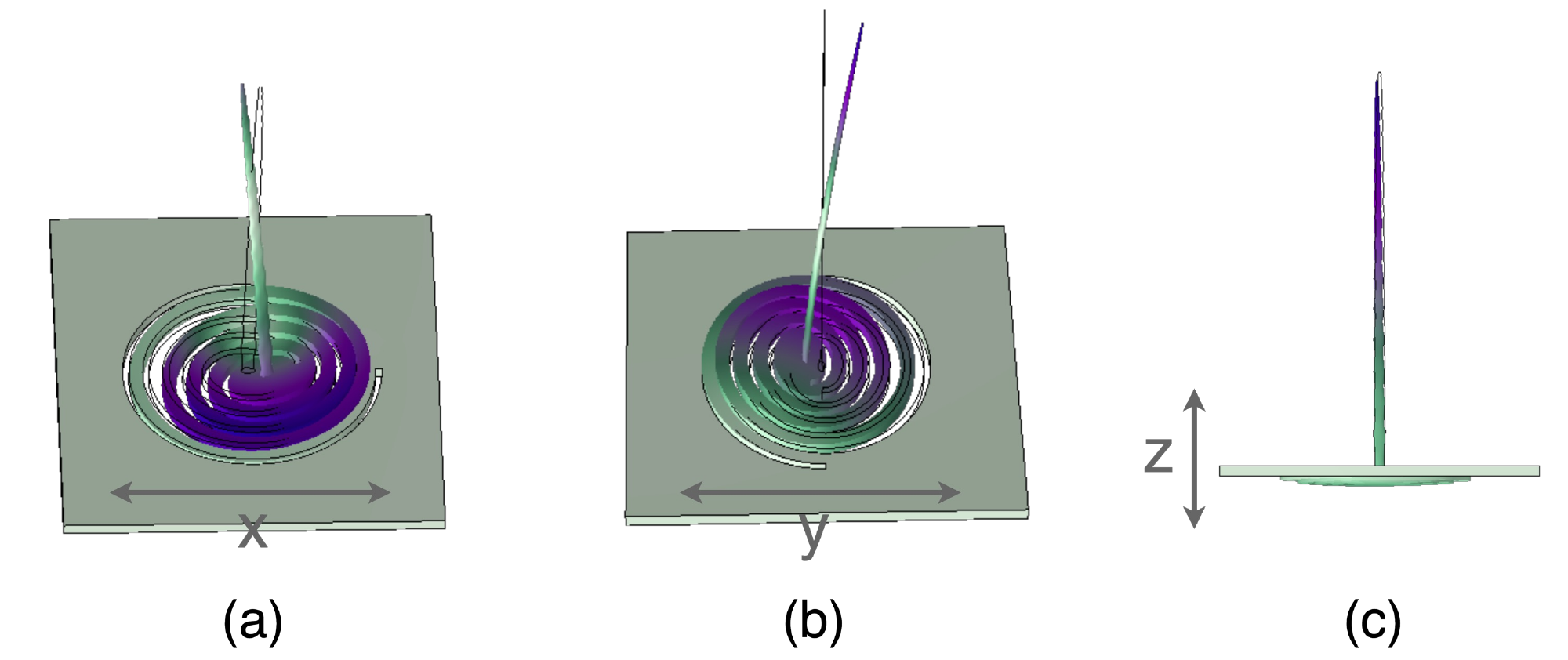}
    \caption{Mode shapes of the spiral-perforated whisker base showing flexible vibrations along (a) the x-axis, (b) the y-axis, and (c) the z-axis.}
    \label{fig:modes}
\end{figure}

The spiral geometry provides multi-directional flexibility, enabling torsional deformation in all three principal directions (x, y, z). As shown in Figure~\ref{fig:modes}a–c, the whisker base exhibits resonant modes in all three directions, confirming its capability for lateral, vertical, and rotational deformation.
For in-plane lateral flexing in the $x$- and $y$-directions, the spiral perforations create space within the whisker base to compress and extend, thereby increasing flexibility in response to horizontal forces and amplifying vibration. In the z-direction, compared to conventional solid bases, which restrict vertical motion due to their resistance to out-of-plane deformation, the spiral-perforated base behaves as a compliant spring, allowing greater vertical displacement of the whisker.
This multi-directional flexibility is crucial for capturing animal wake dynamics, which are inherently three-dimensional~\cite{drucker1999locomotor}. Alternating vortices shed from the animal's tail generate forces in both horizontal directions, while body undulations and vortex interactions produce vertical and torsional components. The spiral-perforated base acts as a multi-axis amplifier, converting these complex 3D vortex forces into measurable vibrations across all directions (x, y, z), thereby enhancing the system's sensitivity to multi-directional vortex-induced oscillations.

\section{Characterization of Spiral Base Parameters on Whisker Vibration Amplification}
\label{sec:char}

We characterize the influence of spiral-shaped whisker base geometry on vibration amplification and frequency response through frequency-response simulations in COMSOL Multiphysics. The vibration amplification is represented by the root mean square (RMS) displacement measured at the sensing point located at the center of the whisker base (Figure~\ref{fig:base_design}b) under harmonic water-pressure excitation at varying frequencies. The geometric parameters investigated include base thickness, number of spiral turns, spiral growth rate, aspect ratio (minor-to-major axis ratio), and perforation slot width. In addition, we analyze the frequency response when vortex forces are applied along different directions (x and y axes). 

The simulation parameters were configured as follows: the rectangular spiral-perforated whisker base was fixed along all four lateral surfaces, and a harmonic water pressure of 15 Pa was applied in the x-direction, aligned with the whisker's major cross-sectional axis. The whisker geometry for characterization consisted of 33 sections, corresponding to lengths of 73.7 mm. The other whisker parameters were $M = 1.3$~mm, $(a,b,k,l) = (0.7, 0.29, 0.62, 0.33)$~mm, and $\alpha = 8.7^\circ$, $\beta = 13.24^\circ$, based on the grey seal whisker geometry reported in previous work
\cite{kamat2024undulating}. The taper factor followed $taper\ factor=(1-n/N)^{0.5}$, where $N$ is the total number of sections and $n$ is the section index ($n=1$ at the whisker base). For the $n$-th cross-section, the elliptical dimensions $(a,b,k,l)$ and inter-section spacing $2M$ were scaled by the corresponding taper factor. The whisker and base material were selected as PLA with density $\rho = 1240 kg/m^3$, Young's modulus $E = 3$GPa, and Poisson's ratio $\nu=0.35$. PLA was selected because it is the primary 3D printing material for fabricating the whisker and base structures. An extra fine mesh was employed in the COMSOL simulations to ensure numerical accuracy. The default whisker base configuration had a thickness of 2 mm, 5 spiral turns, a growth rate of 0.5, an aspect ratio of 0.6, a slot width of 1 mm, with harmonic pressure applied along the x-axis.

\begin{figure}[htbp]
    \centering
    \includegraphics[width=\linewidth]{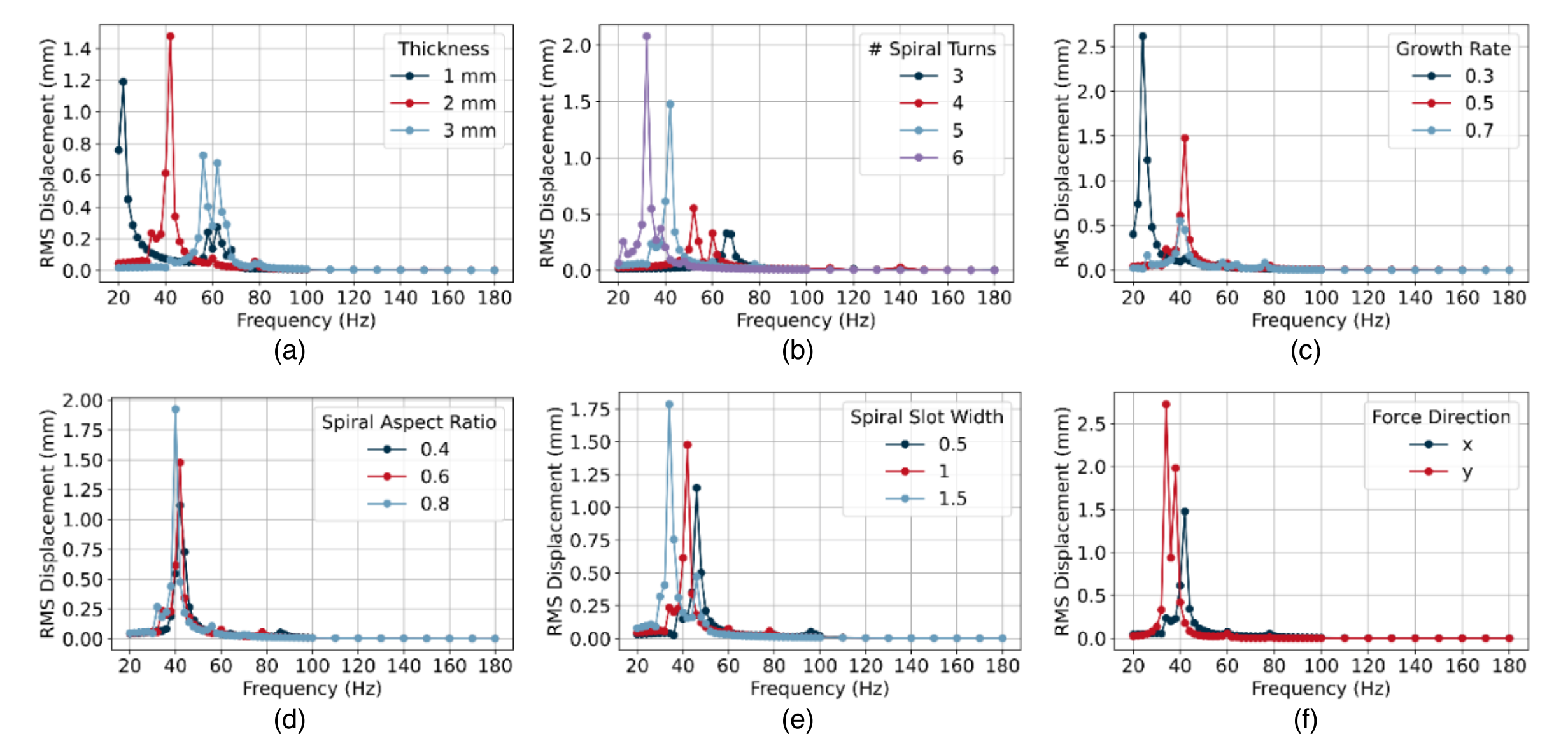}
    \caption{Frequency response of RMS displacement at the sensing location under an applied pressure of 15 Pa. Subplots compare the effects of different whisker base parameters: (a) whisker base thickness, (b) number of spiral turns, (c) spiral growth rate, (d) spiral aspect ratio, (e) spiral slot width, and (f) force direction (x vs. y).}
    \label{fig:characterization}
\end{figure}

The whisker base thickness influences both the amplification frequency range and magnitude. Increasing the base thickness shifts the amplification effect to higher frequencies while reducing its magnitude. As shown in Figure~\ref{fig:characterization}a, when the base thickness increases from 1 mm to 3 mm, the peak frequency of the RMS displacement response shifts from 20 Hz to 60 Hz. This occurs because greater thickness increases the structural stiffness of the whisker base, thereby raising the natural frequency. However, the amplification magnitude at 3 mm thickness is notably smaller than at 1 mm and 2 mm. According to Equation~\ref{eq:x}, increased stiffness $k$ reduces the displacement amplitude under constant force, resulting in lower amplification.

Increasing the number of spiral turns in the whisker base shifts the large displacement frequency range to lower frequencies and enhances the amplification magnitude (see Figure~\ref{fig:characterization}b). This occurs because additional turns increase the effective length of the supporting beam, thereby reducing the structural stiffness. According to Equation~\ref{eq:x}, lower stiffness results in greater amplification at lower frequency ranges.

Increasing the spiral growth rate shifts the sensitive frequency range to higher frequencies while reducing the amplification magnitude. As shown in Figure~\ref{fig:characterization}c, increasing the growth rate from 0.3 to 0.7 raises the peak frequency from 24 Hz to 40 Hz while decreasing the RMS displacement, indicating reduced vibration amplification. This occurs because a higher growth rate increases the effective width of the supporting beam, thereby increasing the structural stiffness and resulting in higher resonant frequencies with lower displacement amplitudes.

The aspect ratio of the spiral has a relatively minor effect on both amplification and frequency range. The aspect ratio is defined as the ratio of the short axis to the long axis of the elliptical spiral. As shown in Figure~\ref{fig:characterization}d, increasing this ratio from 0.4 to 0.8 does not significantly affect the peak frequency or the amplification magnitude. This occurs because the aspect ratio does not substantially alter the dimensions of the supporting beam and therefore has minimal impact on the structural stiffness.

Increasing the spiral slot width shifts the amplification frequency range to lower frequencies while enhancing the amplification magnitude. As shown in Figure~\ref{fig:characterization}e, increasing the slot width from 0.5 mm to 1.5 mm shifts the resonant peak from 42 Hz to 36 Hz, with a corresponding increase in RMS displacement. This occurs because wider slots reduce the effective width of the supporting beam, thereby decreasing the structural stiffness $k$ and resulting in lower resonant frequencies and larger displacements under the same applied force.

Harmonic water pressure applied in the y-direction induces larger vibration displacement at lower frequencies. As shown in Figure~\ref{fig:characterization}f, when the same pressure amplitude (15 Pa) is applied to the whisker surface, the y-direction exhibits greater displacement amplitude. This occurs because the y-axis is perpendicular to the long axis of the elliptical whisker cross-section, resulting in a larger contact area and consequently a greater effective force on the whisker structure. Additionally, due to the elliptical geometry of the whisker base, vibration along the y-direction (corresponding to oscillation along the short axis) represents a lower-order mode with a lower natural frequency. Therefore, the resonant peak occurs at a lower frequency than for the x-direction pressure.

\section{Evaluation of Amplification Effectiveness}

We evaluate the amplification effect of the developed whisker base design through frequency-response simulations in COMSOL Multiphysics. Similar to simulations in characterization, root mean square (RMS) displacement at the sensing point (whisker base center) is used to quantify amplification. We compare the RMS displacement of our design against a baseline whisker structure for three whisker morphologies of different lengths, evaluating its effectiveness across various whisker geometries. Compared to the baseline rectangular-plate support, our design successfully amplifies vortex-induced vibrations for all three whisker morphologies, achieving amplification factors of up to 51.

 We compared the RMS displacement of our whisker base design with the baseline design~\cite{zheng20223d}, in which the whisker was mounted on a single rectangular plate clamped along one short edge, with displacement measured at the opposite free edge. We evaluated three whisker morphologies with 33, 28, and 23 sections, corresponding to lengths of 73.7, 54.2, and 38.3 mm, respectively (referred to as Whisker 1, 2, and 3). These configurations were selected to represent the diverse whisker morphologies and natural frequencies observed in seal whisker arrays~\cite{zheng2025wonders}. The simulation parameter configurations are identical to those in the characterization section.
\begin{figure}[htbp]
    \centering
    \includegraphics[width=\linewidth]{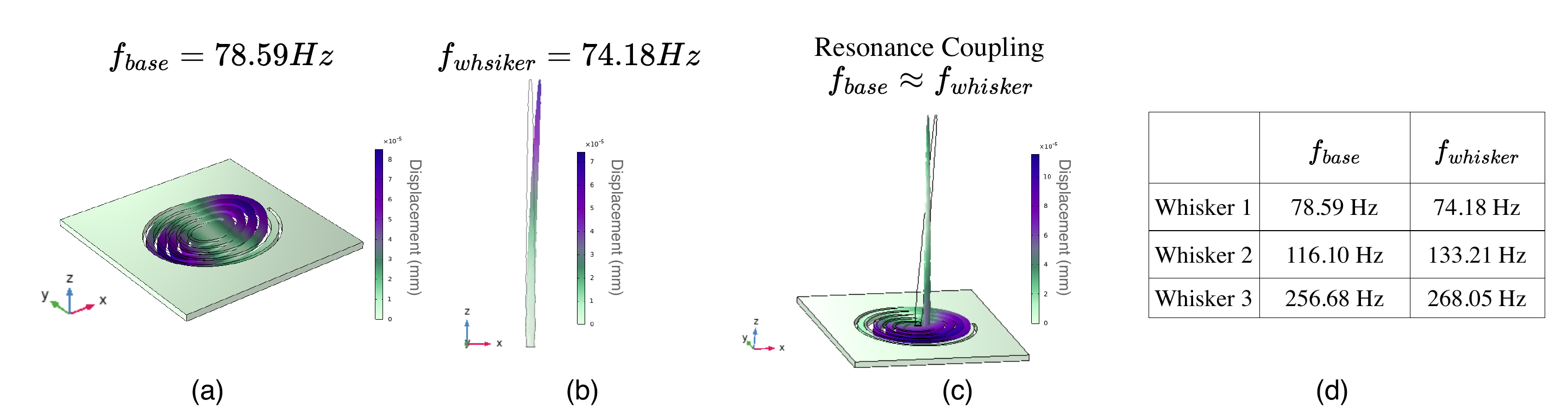}
    \caption{Resonance coupling between whisker base and whisker modes. (a) Base torsional mode shape about the x-axis for Whisker 1 with the natural frequency of 78.59 Hz. (b) Whisker bending mode shape along the x-axis with natural frequency of 74.18 Hz. (c) Coupled system visualization demonstrating near-resonant interaction between base torsion and whisker bending modes. (d) Modal frequency comparison for three whisker geometries showing base ($f_{base}$) and whisker ($f_{whisker}$) natural frequencies in Hz.}
    \label{fig:coupling}
\end{figure}

To achieve resonance coupling, the whisker base geometry was adjusted so that its natural frequency matched the whisker’s natural frequency in the coupled torsional mode. The optimized parameters were: for Whisker 1, thickness = 2 mm and number of turns = 5; for Whisker 2, thickness = 4 mm and number of turns = 5; for Whisker 3, thickness = 2 mm and number of turns = 3. All other geometric parameters remained constant: slot cut width = 1 mm, ellipse major axis = 5 mm, and ellipse minor axis = 3 mm. These configurations were selected to match the whisker's natural frequency with the natural frequency of the whisker base corresponding to the torsional mode shape in the x-direction. Figure~\ref{fig:coupling}a–c illustrates the frequency matching for Whisker 1, showing the base torsional mode, the whisker vibration mode, and the coupled whisker–base system. Figure~\ref{fig:coupling}d summarizes the matching of whisker and base natural frequencies across all three whisker geometries. 

\begin{figure}[htbp]
    \centering
    \includegraphics[width=\linewidth]{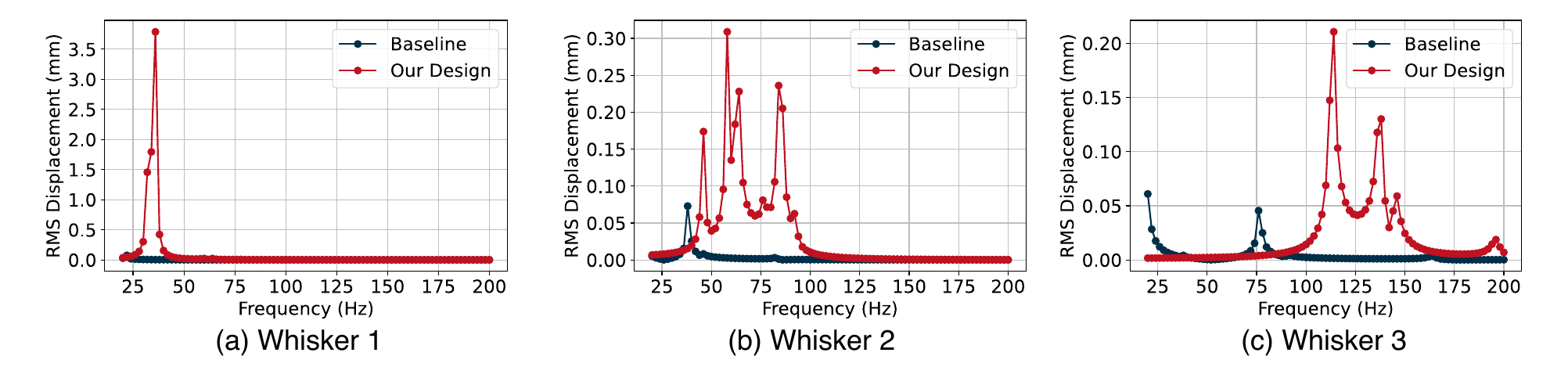}
    \caption{Frequency response of RMS displacement at the center sensing point for (a) Whisker 1 (33 sections), (b) Whisker 2 (28 sections), and (c) Whisker 3 (23 sections), comparing the baseline design and our spiral-perforated whisker base design.}
    \label{fig:results}
\end{figure}

Our spiral-perforated base design achieves significantly larger RMS displacement amplitudes for all three whisker geometries compared to the baseline, demonstrating the effectiveness of the amplification. Figure~\ref{fig:results} shows the frequency response for both our design and the baseline across all three whisker geometries. Whisker 1 exhibits a sensitive frequency range of 30–40 Hz with an amplification factor of 51.13. Whisker 2 is responsive to frequencies of 40–100 Hz with an amplification factor of 3.46, while Whisker 3 demonstrates sensitivity in the 100–150 Hz range with an amplification factor of 4.23. The amplification factor is calculated as the ratio between the maximum RMS displacement among all the frequency values for each design. These sensitive frequency ranges align with the natural frequencies of biological seal whiskers~\cite{zheng2025wonders, zheng20223d}, which primarily span 30–200 Hz. These results confirm that the spiral-perforated whisker base design effectively amplifies vibrations across the diverse frequency ranges characteristic of natural seal whisker arrays.

We observe that the peak frequency of vibration amplification for each whisker is smaller than the natural frequency of the uncoupled whisker or base ($f_{base}$ and $f_{whisker}$). This shift arises because the two systems are fixed together and dynamically coupled. Once coupled, the combined system no longer resonates at the original single-body frequencies. Instead, its modal frequencies split into two: one lower than the lowest original natural frequency and the other higher than the highest~\cite{den1985mechanical}. The lower mode corresponds to the in-phase response, which reinforces motion and produces the amplification peaks we observe~\cite{den1985mechanical,rana1998parametric}. Consequently, the dominant amplification appears at a frequency below either of the original natural frequencies.

\section{Conclusions and Future Works}
In this work, we introduce a seal whisker-inspired sensor for underwater marine animal monitoring and develop a spiral-perforated whisker base design that amplifies weak vibration amplitudes induced by animal-generated wakes. The spiral perforation pattern increases the structural flexibility of the whisker base and reduces effective stiffness, thereby enhancing whisker sensitivity within frequency bands corresponding to target animals. We characterize the influence of geometric parameters on both amplification magnitude and frequency range, providing guidelines for adapting whisker base designs to target frequency bands. To evaluate the amplification effect, we conducted frequency-response simulations on three whisker morphologies targeting 30-40 Hz, 40-100 Hz, and 100-150 Hz, respectively. Our whisker base design achieved up to a 51× increase in RMS displacement compared to the baseline whisker base. The amplification frequency range can be tuned to specific detection targets by adjusting whisker morphology and base geometry.

In the future, we will extend our work on whisker-inspired sensing for underwater marine animal monitoring through real-world experiments. First, we will evaluate the sensor sensitivity and the effectiveness of whisker-inspired sensor design in detecting animal-induced wakes in natural underwater environments. Additionally, we will develop algorithms for detecting, classifying, and tracking marine species to enable long-term, non-invasive monitoring.
Furthermore, we will design whisker arrays comprising sensors tuned to different frequency ranges. By optimizing the cooperative behavior of these arrays, we aim to enhance robustness and adaptability across diverse species and hydrodynamic conditions.

	\section{ACKNOWLEDGEMENTS}
    This work is funded by the Stanford Blume Fellowship.
	
	\section{References}
		\vspace{-1.5ex}
		\bibliographystyle{imac_2021.bst}
		\renewcommand{\refname}{}
		\bibliography{imac}

\begin{thebibliography}{10}

\bibitem{paine1969note}
Paine, R.T.
\newblock ``A note on trophic complexity and community stability''.
\newblock {\em The American Naturalist}, 103(929):91--93 (1969)\removeperiod.

\bibitem{roman2014whales}
Roman, J., Estes, J.A., Morissette, L., Smith, C., Costa, D., McCarthy, J., Nation, J.B., Nicol, S., Pershing, A., and Smetacek, V.
\newblock ``Whales as marine ecosystem engineers''.
\newblock {\em Frontiers in Ecology and the Environment}, 12(7):377--385 (2014)\removeperiod.

\bibitem{heithaus2008predicting}
Heithaus, M.R., Frid, A., Wirsing, A.J., and Worm, B.
\newblock ``Predicting ecological consequences of marine top predator declines''.
\newblock {\em Trends in ecology \& evolution}, 23(4):202--210 (2008)\removeperiod.

\bibitem{dedman2024ecological}
Dedman, S., Moxley, J.H., Papastamatiou, Y.P., Braccini, M., Caselle, J.E., Chapman, D.D., Cinner, J.E., Dillon, E.M., Dulvy, N.K., Dunn, R.E., et~al.
\newblock ``Ecological roles and importance of sharks in the anthropocene ocean''.
\newblock {\em Science}, 385(6708):adl2362 (2024)\removeperiod.

\bibitem{doney2020impacts}
Doney, S.C., Busch, D.S., Cooley, S.R., and Kroeker, K.J.
\newblock ``The impacts of ocean acidification on marine ecosystems and reliant human communities''.
\newblock {\em Annual Review of Environment and Resources}, 45(1):83--112 (2020)\removeperiod.

\bibitem{kroeker2013impacts}
Kroeker, K.J., Kordas, R.L., Crim, R., Hendriks, I.E., Ramajo, L., Singh, G.S., Duarte, C.M., and Gattuso, J.P.
\newblock ``Impacts of ocean acidification on marine organisms: quantifying sensitivities and interaction with warming''.
\newblock {\em Global change biology}, 19(6):1884--1896 (2013)\removeperiod.

\bibitem{alter2024hidden}
Alter, K., Jacquemont, J., Claudet, J., Lattuca, M.E., Barrantes, M.E., Marras, S., Manr{\'\i}quez, P.H., Gonz{\'a}lez, C.P., Fern{\'a}ndez, D.A., Peck, M.A., et~al.
\newblock ``Hidden impacts of ocean warming and acidification on biological responses of marine animals revealed through meta-analysis''.
\newblock {\em Nature Communications}, 15(1):2885 (2024)\removeperiod.

\bibitem{brandt2009harbour}
Brandt, M.J., Diederichs, A., and Nehls, G.
\newblock ``Harbour porpoise responses to pile driving at the horns rev ii offshore wind farm in the danish north sea''.
\newblock {\em Final report to DONG Energy. Husum, Germany, BioConsult SH} (2009)\removeperiod.

\bibitem{tougaard2009pile}
Tougaard, J., Carstensen, J., Teilmann, J., Skov, H., and Rasmussen, P.
\newblock ``Pile driving zone of responsiveness extends beyond 20 km for harbor porpoises (phocoena phocoena (l.))''.
\newblock {\em The Journal of the Acoustical Society of America}, 126(1):11--14 (2009)\removeperiod.

\bibitem{langhamer2012artificial}
Langhamer, O.
\newblock ``Artificial reef effect in relation to offshore renewable energy conversion: state of the art''.
\newblock {\em The Scientific World Journal}, 2012(1):386713 (2012)\removeperiod.

\bibitem{barnette2017potential}
Barnette, M.C.
\newblock ``Potential impacts of artificial reef development on sea turtle conservation in florida'' (2017)\removeperiod.

\bibitem{jepsen2015use}
Jepsen, N., Thorstad, E.B., Havn, T., and Lucas, M.C.
\newblock ``The use of external electronic tags on fish: an evaluation of tag retention and tagging effects''.
\newblock {\em Animal Biotelemetry}, 3(1):49 (2015)\removeperiod.

\bibitem{pine2012design}
Pine, W.E., Hightower, J.E., Coggins, L.G., Lauretta, M.V., and Pollock, K.H.
\newblock ``Design and analysis of tagging studies''.
\newblock {\em Fisheries techniques, 3rd edition. American Fisheries Society, Bethesda, Maryland}, pages 521--572 (2012)\removeperiod.

\bibitem{ruff1995fish}
Ruff, B., Marchant, J., and Frost, A.
\newblock ``Fish sizing and monitoring using a stereo image analysis system applied to fish farming''.
\newblock {\em Aquacultural engineering}, 14(2):155--173 (1995)\removeperiod.

\bibitem{boom2012long}
Boom, B.J., Huang, P.X., Beyan, C., Spampinato, C., Palazzo, S., He, J., Beauxis-Aussalet, E., Lin, S.I., Chou, H.M., Nadarajan, G., et~al.
\newblock ``Long-term underwater camera surveillance for monitoring and analysis of fish populations''.
\newblock In {\em International Workshop on Visual Observation and Analysis of Animal and Insect Behavior (VAIB), in Conjunction with the 21st International Conference on Pattern Recognition, 2012}, pages 1--4 (2012)\removeperiod.

\bibitem{chuang2016underwater}
Chuang, M.C., Hwang, J.N., Ye, J.H., Huang, S.C., and Williams, K.
\newblock ``Underwater fish tracking for moving cameras based on deformable multiple kernels''.
\newblock {\em IEEE Transactions on Systems, Man, and Cybernetics: Systems}, 47(9):2467--2477 (2016)\removeperiod.

\bibitem{hozyn2021review}
Ho{\.z}y{\'n}, S.
\newblock ``A review of underwater mine detection and classification in sonar imagery''.
\newblock {\em Electronics}, 10(23):2943 (2021)\removeperiod.

\bibitem{hodges2011underwater}
Hodges, R.P.
\newblock {\em Underwater acoustics: Analysis, design and performance of sonar}.
\newblock John Wiley \& Sons (2011)\removeperiod.

\bibitem{marage2013sonar}
Marage, J.P. and Mori, Y.
\newblock {\em Sonar and underwater acoustics}.
\newblock John Wiley \& Sons (2013)\removeperiod.

\bibitem{fan2002design}
Fan, Z., Chen, J., Zou, J., Bullen, D., Liu, C., and Delcomyn, F.
\newblock ``Design and fabrication of artificial lateral line flow sensors''.
\newblock {\em Journal of micromechanics and microengineering}, 12(5):655 (2002)\removeperiod.

\bibitem{kottapalli2014touch}
Kottapalli, A.G.P., Asadnia, M., Miao, J., and Triantafyllou, M.
\newblock ``Touch at a distance sensing: lateral-line inspired mems flow sensors''.
\newblock {\em Bioinspiration \& biomimetics}, 9(4):046011 (2014)\removeperiod.

\bibitem{venturelli2012hydrodynamic}
Venturelli, R., Akanyeti, O., Visentin, F., Je{\v{z}}ov, J., Chambers, L.D., Toming, G., Brown, J., Kruusmaa, M., Megill, W.M., and Fiorini, P.
\newblock ``Hydrodynamic pressure sensing with an artificial lateral line in steady and unsteady flows''.
\newblock {\em Bioinspiration \& biomimetics}, 7(3):036004 (2012)\removeperiod.

\bibitem{chambers2014fish}
Chambers, L.D., Akanyeti, O., Venturelli, R., Je{\v{z}}ov, J., Brown, J., Kruusmaa, M., Fiorini, P., and Megill, W.
\newblock ``A fish perspective: detecting flow features while moving using an artificial lateral line in steady and unsteady flow''.
\newblock {\em Journal of The Royal Society Interface}, 11(99):20140467 (2014)\removeperiod.

\bibitem{liu2016review}
Liu, G., Wang, A., Wang, X., and Liu, P.
\newblock ``A review of artificial lateral line in sensor fabrication and bionic applications for robot fish''.
\newblock {\em Applied bionics and biomechanics}, 2016(1):4732703 (2016)\removeperiod.

\bibitem{zheng2021creating}
Zheng, X., Kamat, A.M., Cao, M., and Kottapalli, A.G.P.
\newblock ``Creating underwater vision through wavy whiskers: A review of the flow-sensing mechanisms and biomimetic potential of seal whiskers''.
\newblock {\em Journal of the Royal Society Interface}, 18(183):20210629 (2021)\removeperiod.

\bibitem{dehnhardt2001hydrodynamic}
Dehnhardt, G., Mauck, B., Hanke, W., and Bleckmann, H.
\newblock ``Hydrodynamic trail-following in harbor seals (phoca vitulina)''.
\newblock {\em Science}, 293(5527):102--104 (2001)\removeperiod.

\bibitem{hanke2010harbor}
Hanke, W., Witte, M., Miersch, L., Brede, M., Oeffner, J., Michael, M., Hanke, F., Leder, A., and Dehnhardt, G.
\newblock ``Harbor seal vibrissa morphology suppresses vortex-induced vibrations''.
\newblock {\em Journal of Experimental Biology}, 213(15):2665--2672 (2010)\removeperiod.

\bibitem{zhang2021harbor}
Zhang, X., Shan, X., Xie, T., Miao, J., Du, H., and Song, R.
\newblock ``Harbor seal whisker inspired self-powered piezoelectric sensor for detecting the underwater flow angle of attack and velocity''.
\newblock {\em Measurement}, 172:108866 (2021)\removeperiod.

\bibitem{dai2024biomimetic}
Dai, H., Zhang, C., Hu, H., Hu, Z., Sun, H., Liu, K., Li, T., Fu, J., Zhao, P., and Yang, H.
\newblock ``Biomimetic hydrodynamic sensor with whisker array architecture and multidirectional perception ability''.
\newblock {\em Advanced Science}, 11(38):2405276 (2024)\removeperiod.

\bibitem{zheng20223d}
Zheng, X., Kamat, A.M., Krushynska, A.O., Cao, M., and Kottapalli, A.G.P.
\newblock ``3d printed graphene piezoresistive microelectromechanical system sensors to explain the ultrasensitive wake tracking of wavy seal whiskers''.
\newblock {\em Advanced Functional Materials}, 32(47):2207274 (2022)\removeperiod.

\bibitem{beem2012calibration}
Beem, H., Hildner, M., and Triantafyllou, M.
\newblock ``Calibration and validation of a harbor seal whisker-inspired flow sensor''.
\newblock {\em Smart Materials and Structures}, 22(1):014012 (2012)\removeperiod.

\bibitem{kottapalli2015harbor}
Kottapalli, A.G.P., Asadnia, M., Miao, J., and Triantafyllou, M.S.
\newblock ``Harbor seal whisker inspired flow sensors to reduce vortex-induced vibrations''.
\newblock In {\em 2015 28th IEEE International Conference on Micro Electro Mechanical Systems (MEMS)}, pages 889--892. IEEE (2015)\removeperiod.

\bibitem{nauen2002quantification}
Nauen, J.C. and Lauder, G.V.
\newblock ``Quantification of the wake of rainbow trout (oncorhynchus mykiss) using three-dimensional stereoscopic digital particle image velocimetry''.
\newblock {\em Journal of Experimental Biology}, 205(21):3271--3279 (2002)\removeperiod.

\bibitem{drucker2001wake}
Drucker, E.G. and Lauder, G.V.
\newblock ``Wake dynamics and fluid forces of turning maneuvers in sunfish''.
\newblock {\em Journal of Experimental Biology}, 204(3):431--442 (2001)\removeperiod.

\bibitem{yaul2012flexible}
Yaul, F.M., Bulovic, V., and Lang, J.H.
\newblock ``A flexible underwater pressure sensor array using a conductive elastomer strain gauge''.
\newblock {\em Journal of microelectromechanical systems}, 21(4):897--907 (2012)\removeperiod.

\bibitem{hanke2004hydrodynamic}
Hanke, W. and Bleckmann, H.
\newblock ``The hydrodynamic trails of lepomis gibbosus (centrarchidae), colomesus psittacus (tetraodontidae) and thysochromis ansorgii (cichlidae) investigated with scanning particle image velocimetry''.
\newblock {\em Journal of Experimental Biology}, 207(9):1585--1596 (2004)\removeperiod.

\bibitem{kamat2024undulating}
Kamat, A.M., Zheng, X., Bos, J., Cao, M., Triantafyllou, M.S., and Kottapalli, A.G.P.
\newblock ``Undulating seal whiskers evolved optimal wavelength-to-diameter ratio for efficient reduction in vortex-induced vibrations''.
\newblock {\em Advanced Science}, 11(2):2304304 (2024)\removeperiod.

\bibitem{triantafyllou2000hydrodynamics}
Triantafyllou, M.S., Triantafyllou, G., and Yue, D.K.
\newblock ``Hydrodynamics of fishlike swimming''.
\newblock {\em Annual review of fluid mechanics}, 32(1):33--53 (2000)\removeperiod.

\bibitem{tytell2004hydrodynamics}
Tytell, E.D. and Lauder, G.V.
\newblock ``The hydrodynamics of eel swimming: I. wake structure''.
\newblock {\em Journal of Experimental Biology}, 207(11):1825--1841 (2004)\removeperiod.

\bibitem{sfakiotakis2002review}
Sfakiotakis, M., Lane, D.M., and Davies, J.B.C.
\newblock ``Review of fish swimming modes for aquatic locomotion''.
\newblock {\em IEEE Journal of oceanic engineering}, 24(2):237--252 (2002)\removeperiod.

\bibitem{wieskotten2011hydrodynamic}
Wieskotten, S., Mauck, B., Miersch, L., Dehnhardt, G., and Hanke, W.
\newblock ``Hydrodynamic discrimination of wakes caused by objects of different size or shape in a harbour seal (phoca vitulina)''.
\newblock {\em Journal of Experimental Biology}, 214(11):1922--1930 (2011)\removeperiod.

\bibitem{zheng2025wonders}
Zheng, X., Kamat, A.M., Cao, M., Triantafyllou, M.S., and Kottapalli, A.G.P.
\newblock ``Wonders of harbor and grey seal whiskers: Morphology, natural frequencies, and 3d modeling''.
\newblock {\em Advanced Science}, 12(23):2500724 (2025)\removeperiod.

\bibitem{morkovin1964flow}
Morkovin, M.
\newblock ``Flow around circular cylinder-kaleidoscope of challenging fluid phenomena''.
\newblock In {\em Proc. Symp. Fully Separated Flows, Philadelphia, 1964}, pages 102--118. ASME (1964)\removeperiod.

\bibitem{beaudan1995numerical}
Beaudan, P.B.
\newblock {\em Numerical experiments on the flow past a circular cylinder at sub-critical Reynolds number}.
\newblock Stanford University (1995)\removeperiod.

\bibitem{den1985mechanical}
Den~Hartog, J.P.
\newblock {\em Mechanical vibrations}.
\newblock Courier Corporation (1985)\removeperiod.

\bibitem{rana1998parametric}
Rana, R. and Soong, T.T.
\newblock ``Parametric study and simplified design of tuned mass dampers''.
\newblock {\em Engineering structures}, 20(3):193--204 (1998)\removeperiod.

\bibitem{drucker1999locomotor}
Drucker, E.G. and Lauder, G.V.
\newblock ``Locomotor forces on a swimming fish: three-dimensional vortex wake dynamics quantified using digital particle image velocimetry''.
\newblock {\em Journal of Experimental Biology}, 202(18):2393--2412 (1999)\removeperiod.

\end{thebibliography}

\end{document}